\PassOptionsToPackage{unicode}{hyperref}
\PassOptionsToPackage{hyphens}{url}
\documentclass[
]{article}
\usepackage{lmodern}
\usepackage{amssymb,amsmath}
\usepackage{ifxetex,ifluatex}
\ifnum 0\ifxetex 1\fi\ifluatex 1\fi=0 % if pdftex
  \usepackage[T1]{fontenc}
  \usepackage[utf8]{inputenc}
  \usepackage{textcomp} % provide euro and other symbols
\else % if luatex or xetex
  \usepackage{unicode-math}
  \defaultfontfeatures{Scale=MatchLowercase}
  \defaultfontfeatures[\rmfamily]{Ligatures=TeX,Scale=1}
\fi
\IfFileExists{upquote.sty}{\usepackage{upquote}}{}
\IfFileExists{microtype.sty}{% use microtype if available
  \usepackage[]{microtype}
  \UseMicrotypeSet[protrusion]{basicmath} % disable protrusion for tt fonts
}{}
\makeatletter
\@ifundefined{KOMAClassName}{% if non-KOMA class
  \IfFileExists{parskip.sty}{%
    \usepackage{parskip}
  }{% else
    \setlength{\parindent}{0pt}
    \setlength{\parskip}{6pt plus 2pt minus 1pt}}
}{% if KOMA class
  \KOMAoptions{parskip=half}}
\makeatother
\usepackage{xcolor}
\IfFileExists{xurl.sty}{\usepackage{xurl}}{} % add URL line breaks if available
\IfFileExists{bookmark.sty}{\usepackage{bookmark}}{\usepackage{hyperref}}
\hypersetup{
  pdftitle={Analysing Social Media Network Data with R},
  pdfauthor={a Working Paper and Tutorial by Dennis Klinkhammer},
  hidelinks,
  pdfcreator={LaTeX via pandoc}}
\usepackage[margin=1in]{geometry}
\usepackage{color}
\usepackage{fancyvrb}

\DefineVerbatimEnvironment{Highlighting}{Verbatim}{commandchars=\\\{\}}
\usepackage{framed}
\definecolor{shadecolor}{RGB}{248,248,248}
\newenvironment{Shaded}{\begin{snugshade}}{\end{snugshade}}

\newcommand{\DataTypeTok}[1]{\textcolor[rgb]{0.13,0.29,0.53}{#1}}
\newcommand{\DecValTok}[1]{\textcolor[rgb]{0.00,0.00,0.81}{#1}}

\newcommand{\FloatTok}[1]{\textcolor[rgb]{0.00,0.00,0.81}{#1}}

\newcommand{\KeywordTok}[1]{\textcolor[rgb]{0.13,0.29,0.53}{\textbf{#1}}}
\newcommand{\NormalTok}[1]{#1}
\newcommand{\OperatorTok}[1]{\textcolor[rgb]{0.81,0.36,0.00}{\textbf{#1}}}
\newcommand{\OtherTok}[1]{\textcolor[rgb]{0.56,0.35,0.01}{#1}}

\newcommand{\StringTok}[1]{\textcolor[rgb]{0.31,0.60,0.02}{#1}}

\usepackage{graphicx,grffile}
\makeatletter
\def\maxwidth{\ifdim\Gin@nat@width>\linewidth\linewidth\else\Gin@nat@width\fi}
\def\maxheight{\ifdim\Gin@nat@height>\textheight\textheight\else\Gin@nat@height\fi}
\makeatother
\setkeys{Gin}{width=\maxwidth,height=\maxheight,keepaspectratio}
\makeatletter
\def\fps@figure{htbp}
\makeatother
\title{Analysing Social Media Network Data with R}
\usepackage{etoolbox}
\makeatletter
\providecommand{\subtitle}[1]{% add subtitle to \maketitle
  \apptocmd{\@title}{\par {\large #1 \par}}{}{}
}
\makeatother
\subtitle{Semi-Automated Screening of Users, Comments and Communication Patterns}
\author{a Working Paper and Tutorial by Dennis Klinkhammer}
\date{}

\begin{document}
\maketitle
\begin{abstract}
Communication on social media platforms is not only culturally and
politically relevant, it is also increasingly widespread across
societies. Users not only communicate via social media platforms, but
also search specifically for information, disseminate it or post
information themselves. However, fake news, hate speech and even
radicalizing elements are part of this modern form of communication:
Sometimes with far-reaching effects on individuals and societies. A
basic understanding of these mechanisms and communication patterns could
help to counteract negative forms of communication, e.g.~bullying among
children or extreme political points of view. To this end, a method will
be presented in order to break down the underlying communication
patterns, to trace individual users and to inspect their comments and
range on social media platforms; Or to contrast them later on via
qualitative research. This approeach can identify particularly active
users with an accuracy of 100 percent, if the framing social networks as
well as the topics are taken into account. However, methodological as
well as counteracting approaches must be even more dynamic and flexible
to ensure sensitivity and specifity regarding users who spread hate
speech, fake news and radicalizing elements.
\end{abstract}

\hypertarget{keywords}{%
\paragraph{Keywords}\label{keywords}}

Social Media Analysis, Social Networks, Communication Patterns, Network
Data Screening

\hypertarget{relevance-of-social-media-networks}{%
\subsection{Relevance of Social Media
Networks}\label{relevance-of-social-media-networks}}

Forecasts assume that in the year 2021 the number of contributions of
active social media users will increase to 3.02 billion per month
worldwide (Pereira-Kohatsu et al.~2019). The importance of such
contributions for forming a political opinion was especially set in
scene by the 45th President of the United States of America, Donald
Trump. He used social media platforms frequently for sharing and
commenting his political actions; Sometimes by using aggressive,
inappropriate, misleading, racist and / or otherwise disparaging
comments (Ott 2016). In addition, social media platforms can shape the
cultural habitus of parts of societies and contribute to the transfer of
knowledge.

But fake news, hate speech as well as radicalizing elements show an
increasing number on social media platforms, as well (Reichelmann et
al.~2020). However, such social media elements - often on seemingly
harmless social media platforms like YouTube and accessible to children
(Machackova et al.~2020) - can also be used to investigate
communications patterns inside social networks, to focus the role of
individual users and what influence the content of their comments and
actions might have on the underlying structures of a social network.
Giving as well as withholding likes and replies on social media
platforms, for example, are actions that can be analysed on a
quantitative basis. These are individual phenomena occurring en masse,
of which some seem more relevant than others when it comes to generating
and distributing fake news, hate speech and radicalizing elements.

YouTube, for example, had 2.1 billion active users in 2020 and
represents one of the most frequently used social media platforms
worldwide (Clement 2020). Along with videos on cultural, political and
even private content, there is a large number of comments, likes and
replies. Some of these comments include fake news, hate speech and
radicalizing elements, that need to be assessed critically (Awan 2017).
However, focusing individual aspects of social networks manually,
e.g.~for getting deeper insights or contrasting analyses, tends to be
very time-consuming for researchers from various scientific disciplines.
Both, qualitative and quantitative analyses, are faced with a very large
and continuously growing number of users and contributions; Each of them
bringing an influential dynamic that needs to be considered in every
methodological approach.

Therefore, this contribution presents how singular social media
contributions can be identified throughout a mass of contributions, how
individual users and their activities can be traced and, furthermore,
how important it is to consider different communication patterns,
depending on the content. One of the results is a graphic representation
of individual users and the range of their contributions on social media
platforms. Another result can be considered the basis for creating
automated screening methods - in addition to this semi-automated
approach - and how to generate qualitative datasets out of social media
data for further research.

This contribution includes all software requirements, a full disclosure
of codes for the R programming language, the entire process of accessing
and pre-precessing social media based datasets and how to analyse and
visualize the included variables. As a result, this working paper is
mainly supposed to be a methodological tutorial for students,
researchers and users of social media platforms.

\hypertarget{software-requirements}{%
\subsection{Software Requirements}\label{software-requirements}}

This working paper is written in \emph{R Markdown}, so it uses the R
programming language. A free software environment for using R is
available for Linux, macOS by Apple and Windows by Microsoft. The main
purpose of R is statistical computing and it is both used for manual
quantitative analyses as well as automated or semi-automated analyses.
When it comes to Big Data, R can also be used for unsupervised and
supervised Machine Learning.

In detail, R is an object based programming language. Therefore,
datasets, variables, cases, values as well as functions can be applied
as a combination of objects, as it is necessary for Analysing Social
Media Network Data. All commands, as combinations of functions,
datasets, variables, cases, values and functions will be highlighted
within this working paper, so that they can be used as step by step
tutorial. The commands have to be entered directly into the R terminal,
which is available after downloading, installing and starting the R
software environment.

\hypertarget{preparations---1-attaching-necessary-packages}{%
\subsection{Preparations - 1: Attaching necessary
Packages}\label{preparations---1-attaching-necessary-packages}}

The Analysis of Social Media Network Data requires four additional
packages in order to expand the range of basic R functions. All packages
can be installed by using the \emph{install(\ldots)} command and can be
attached via the \emph{library(\ldots)} command by typing the following
commands directly into the R terminal.

Since the Analysis of Social Media Network Data can be described as a
huge process regarding the underlying data structure, it is necessary to
break down this process and its underlying data structure into
manageable little pieces. A package that is specifically designed to do
so is called \emph{dplyr}. It can split, apply and combine data for
further analysis (Wickham 2020). The package \emph{dplyr} can be
installed and attached as follows:

\begin{Shaded}
\begin{Highlighting}[]
\KeywordTok{install.packages}\NormalTok{(}\StringTok{"plyr"}\NormalTok{, }\DataTypeTok{dependencies=}\OtherTok{TRUE}\NormalTok{)}
\KeywordTok{library}\NormalTok{(plyr)}
\end{Highlighting}
\end{Shaded}

Variables regarding the social network structures on social media
platforms can easily be accessed by using the \emph{vosonSML} package.
It expands the range of basic R functions in order to collect data from
popular social media platforms like YouTube, Twitter or Reddit (Graham
et al.~2020). This package will provide access to comments, likes,
replies and additional information regarding the social network
structures on these social media platforms. These commands will install
and attach the \emph{vosonSML} package:

\begin{Shaded}
\begin{Highlighting}[]
\KeywordTok{install.packages}\NormalTok{(}\StringTok{"vosonSML"}\NormalTok{, }\DataTypeTok{dependencies=}\OtherTok{TRUE}\NormalTok{)}
\KeywordTok{library}\NormalTok{(vosonSML)}
\end{Highlighting}
\end{Shaded}

For keeping the necessary R commands tidy and structured, the package
\emph{magrittr} can be used. It provides a forward-pipe operator that
allows for chaining several commands (Bache/Wickham 2020). As a result,
complementary commands for a single analysis step can be called up
together. The package \emph{magrittr} can also be installed and attached
by using the \emph{install(\ldots)} and \emph{library(\ldots)} commands:

\begin{Shaded}
\begin{Highlighting}[]
\KeywordTok{install.packages}\NormalTok{(}\StringTok{"magrittr"}\NormalTok{, }\DataTypeTok{dependencies=}\OtherTok{TRUE}\NormalTok{)}
\KeywordTok{library}\NormalTok{(magrittr)}
\end{Highlighting}
\end{Shaded}

The final package is called \emph{stringr}. Since text, like in social
media comments, is represented by character variables in R, a package
that can process and - if necessary - manipulate individual characters
within the strings of a character variable is required (Wickham 2019);
And a string is marked either by single quote signs or double quote
signs. In order to install and attach the \emph{stringr} package,
following commands can be typed in the R terminal:

\begin{Shaded}
\begin{Highlighting}[]
\KeywordTok{install.packages}\NormalTok{(}\StringTok{"stringr"}\NormalTok{, }\DataTypeTok{dependencies=}\OtherTok{TRUE}\NormalTok{)}
\KeywordTok{library}\NormalTok{(stringr)}
\end{Highlighting}
\end{Shaded}

Additional note: It is possible that further packages have to be
installed in the R environment. This will be automatically checked via
the extension of the \emph{install (\ldots)} command with
\emph{dependencies = TRUE} that will install additional packages, if
necessary. In a freshly created R environment (based upon version 4.0.3
of R), the four packages listed above have been running sufficient on
Ubuntu 20.04 LTS (Linux Kernel 5.9.11), macOS Big Sur (11.0) and Windows
10 (20H2), each brought to application on RStudio.

\hypertarget{preparations---2-requesting-an-application-programming-interface}{%
\subsection{Preparations - 2: Requesting an Application Programming
Interface}\label{preparations---2-requesting-an-application-programming-interface}}

In order to perform an Analysis of Social Media Network Data, analytical
access to social media platforms and its underlying variables is
required. This can be achieved via an Application Programming Interface
(API). An API is a set of functions and procedures allowing the creation
of applications that access the features or data of an operating system,
application, or other services like social media platforms (Fielding
2000). In short: It lets the \emph{vosonSML} package in R communicate
with YouTube, Twitter and Reddit as social media platforms.

For scientific purpose APIs can be used to generate datasets with a
variety of variables regarding the users of social media platforms and
their behavior on these platforms. How an API can be included in the R
commands for performing an Analysis of Social Media Network Data is
presented below. This is an example for the social media platform
YouTube, that can be easily adapted for similar application with Twitter
and Reddit:

\begin{Shaded}
\begin{Highlighting}[]
\NormalTok{my_apiKeyYoutube<-}\StringTok{"ENTER YOUR APPLICATION PROGRAMMING INTERFACE HERE"}
\NormalTok{apiKeyYoutube<-}\KeywordTok{Authenticate}\NormalTok{(}\StringTok{"youtube"}\NormalTok{, }\DataTypeTok{apiKey=}\NormalTok{my_apiKeyYoutube)}
\end{Highlighting}
\end{Shaded}

Since YouTube belongs to Google, a Google Developers Account is
necessary in order to request an API. Each API will be assigned to a
specific project, such as e.g.~``Social Media Analysis''. Creating an
account, logging in and starting a new project with an specific API is
possible within the Google Developers console:
\emph{\url{https://console.developers.google.com/project/}}.

There are similar procedures for requesting an API for Twitter or
Reddit. Twitter offers a developers program, similar to YouTube; Reddit
provides an \emph{wiki} on requesting an API.

\hypertarget{accessing-data---1-selecting-content}{%
\subsection{Accessing Data - 1: Selecting
Content}\label{accessing-data---1-selecting-content}}

Once a social media platform has been selected for an Analysis of Social
Media Network Data, an access point must be determined. In this context,
for example, access points can be relevant content on social media
platforms. Sticking to the example with YouTube, one would not only pick
a video as access point, but one with content relevant to current
affairs, created by a specific person or on a specific topic. In most
scientific cases, the research interest will determine the relevance of
the access point.

Each access point can be directly addressed via an unique identification
number, which is clearly assigned to each content on social media
platforms. On YouTube these identification numbers are combinations of
letters and digits at the end of each link to a video, separated by an
equals sign. The combination of letters from a-z, A-Z and digits from
0-9 provides a unique identification number than can be entered in the R
terminal as follows:

\begin{Shaded}
\begin{Highlighting}[]
\NormalTok{videoIDs=}\KeywordTok{c}\NormalTok{(}\StringTok{"ENTER IDENTIFICATION NUMBER OF SOCIAL MEDIA CONTENT HERE"}\NormalTok{)}
\end{Highlighting}
\end{Shaded}

Referring to the presidential election 2020 in the U.S. for example,
YouTube provides a news video called \emph{US Elections: Trump's
Statement in Full - BBC News}, identified by the unique identification
number \emph{KtNZV7qezMM}. This news video has been watched more than
200.000 times within two days and several users have generated comments,
likes and replies below this video, which reflects a moment of current
affairs. The Analysis of Social Media Network Data provides methods for
an quick identification of particularly active users, their comments and
the underlying communication patterns within that social media content.

Twitter and Reddit provide similar access points and it is possible to
combine several access points, separated by a comma between the unique
identification numbers within the above presented R command.

\hypertarget{accessing-data---2-generating-a-dataset}{%
\subsection{Accessing Data - 2: Generating a
Dataset}\label{accessing-data---2-generating-a-dataset}}

This step uses the \emph{Collect(\ldots)} command of the \emph{vosonSML}
package and refers to the generated objects \emph{apiKeyYoutube} and
\emph{videoIDs} of the previous steps in order to access YouTube. In
addition, the commands \emph{writeToFile=TRUE} and \emph{verbose=TRUE}
not only allow to write the collected data to a file, but also to
collect additional information on the process of data gathering, if
necessary. Furthermore, the maximum number of collected comments is most
relevant to the Analysis of Social Media Network Data, since the size
could influence the results, e.g.~when the relevance of a topic changes
over a period of time or the public opinion tends to change. This
example uses the command \emph{maxComments=200} and therefore collects
the latest 200 comments and all replies to these comments, without
counting the replies as one of the latest 200 comments. As a result,
more than 200 comments can be collected by typing the following command
into the R terminal:

\begin{Shaded}
\begin{Highlighting}[]
\NormalTok{myYoutubeData<-}\KeywordTok{Collect}\NormalTok{(apiKeyYoutube, videoIDs,}
  \DataTypeTok{writeToFile=}\OtherTok{TRUE}\NormalTok{, }\DataTypeTok{verbose=}\OtherTok{TRUE}\NormalTok{, }\DataTypeTok{maxComments=}\DecValTok{200}\NormalTok{)}
\end{Highlighting}
\end{Shaded}

Hence, a new dataset \emph{myYoutubeData} will be generated. The
\emph{str(\ldots)} command and \emph{head(\ldots)} command would provide
first insights into this initial dataset, but before doing so the
dataset will be pre-processed in order to get a tidy dataset.

\hypertarget{data-pre-processing---1-inspecting-the-variables}{%
\subsection{Data Pre-Processing - 1: Inspecting the
Variables}\label{data-pre-processing---1-inspecting-the-variables}}

Data pre-processing is used to check datasets for irrelevant and
redundant information present or noisy and unreliable data. Targeted
research questions with a firm theoretical background also allow
focusing the most relevant variables. For an thorough Analysis of Social
Media Network Data the authors of social media comments would be
relevant, the sentences they have written, how often someone replied to
their comments and how many likes they got for their comments. Focusing
the relevant variables provides a tidy dataset as result of data
pre-processing. The variables mentioned can be transferred to a first
tidy dataset to start with and be labeled accordingly by using the
following command:

\begin{Shaded}
\begin{Highlighting}[]
\NormalTok{easy_dataset <-}\StringTok{ }\KeywordTok{data.frame}\NormalTok{(myYoutubeData}\OperatorTok{$}\NormalTok{AuthorDisplayName,}
\NormalTok{  myYoutubeData}\OperatorTok{$}\NormalTok{Comment, myYoutubeData}\OperatorTok{$}\NormalTok{ReplyCount, myYoutubeData}\OperatorTok{$}\NormalTok{LikeCount)}
\KeywordTok{colnames}\NormalTok{(easy_dataset) <-}\StringTok{ }\KeywordTok{c}\NormalTok{(}\StringTok{"author"}\NormalTok{, }\StringTok{"sentence"}\NormalTok{, }\StringTok{"replies"}\NormalTok{, }\StringTok{"likes"}\NormalTok{)}
\end{Highlighting}
\end{Shaded}

This command leads to a dataset called \emph{easy\_dataset} with four
variables: \emph{author}, \emph{sentence}, \emph{replies} and
\emph{likes}. A first insight can be called via the
\emph{summary(\ldots)} command applied on the \emph{easy\_dataset} in
the R terminal:

\begin{Shaded}
\begin{Highlighting}[]
\KeywordTok{summary}\NormalTok{(easy_dataset)}
\end{Highlighting}
\end{Shaded}

The output should look like this in order to focus the number of cases
and the type of the variables:

\begin{verbatim}
##     author            sentence           replies             likes          
##  Length:252         Length:252         Length:252         Length:252        
##  Class :character   Class :character   Class :character   Class :character  
##  Mode  :character   Mode  :character   Mode  :character   Mode  :character
\end{verbatim}

As a result of this command, the four variables and the type of these
variables are shown. All four variables are currently shown as character
variables. However, this would only be true if all variables included
letters. The names of the authors of the comments as well as their
written sentences will most likely include letters, so the variables
\emph{author} and \emph{sentence} should remain character variables; At
least for the moment. Since the variables \emph{replies} and
\emph{likes} contain numbers and not letters, simply by counting the
number of replies and likes, they should therefore be converted into
numeric variables:

\begin{Shaded}
\begin{Highlighting}[]
\NormalTok{easy_dataset}\OperatorTok{$}\NormalTok{likes <-}\StringTok{ }\KeywordTok{as.numeric}\NormalTok{(}\KeywordTok{as.character}\NormalTok{(easy_dataset}\OperatorTok{$}\NormalTok{likes))}
\NormalTok{easy_dataset}\OperatorTok{$}\NormalTok{replies <-}\StringTok{ }\KeywordTok{as.numeric}\NormalTok{(}\KeywordTok{as.character}\NormalTok{(easy_dataset}\OperatorTok{$}\NormalTok{replies))}
\NormalTok{easy_dataset}\OperatorTok{$}\NormalTok{sentence <-}\StringTok{ }\KeywordTok{as.character}\NormalTok{(easy_dataset}\OperatorTok{$}\NormalTok{sentence)}
\NormalTok{easy_dataset}\OperatorTok{$}\NormalTok{author <-}\StringTok{ }\KeywordTok{as.character}\NormalTok{(easy_dataset}\OperatorTok{$}\NormalTok{author)}
\end{Highlighting}
\end{Shaded}

The results of this procedure can be checked by repeating the
\emph{summary(\ldots)} command in the R terminal:

\begin{Shaded}
\begin{Highlighting}[]
\KeywordTok{summary}\NormalTok{(easy_dataset)}
\end{Highlighting}
\end{Shaded}

However, this output is not only a snapshot and may have developed
further in a later step of this analysis, but may also vary due to the
parameter in the \emph{maxComments} setting while generating the initial
dataset:

\begin{verbatim}
##     author            sentence            replies            likes       
##  Length:252         Length:252         Min.   : 0.0000   Min.   : 0.000  
##  Class :character   Class :character   1st Qu.: 0.0000   1st Qu.: 0.000  
##  Mode  :character   Mode  :character   Median : 0.0000   Median : 0.000  
##                                        Mean   : 0.2063   Mean   : 1.401  
##                                        3rd Qu.: 0.0000   3rd Qu.: 1.000  
##                                        Max.   :10.0000   Max.   :52.000
\end{verbatim}

Now the minimum, mean and maximum number of replies and likes for the
variables \emph{replies} and \emph{likes} can be taken directly from the
output and they can be interpreted as numeric variables. For example,
there seems to be at least one author of social media comments who got a
maximum of \emph{10} replies and at least another one - or the same one,
since that can not be determined before further analysis - got a maximum
of \emph{52} likes.

There is also a way for generating numeric variables on basis of
character variables. However, this requires a methodical approach on
dealing with duplicate cases. Therefore, in the next section, a
corresponding procedure is presented without affecting the informative
value of the dataset by focusing the cases.

\hypertarget{data-pre-processing---2-inspecting-the-cases}{%
\subsection{Data Pre-Processing - 2: Inspecting the
Cases}\label{data-pre-processing---2-inspecting-the-cases}}

By counting the number of comments of an user via the variable
\emph{author} a new variable \emph{comments} can be generated. Same
procedure is applicable on the number of words within the variable
\emph{sentence}. Numeric variables on basis of character variables can
be generated like this:

\begin{Shaded}
\begin{Highlighting}[]
\NormalTok{comments <-}\StringTok{ }\KeywordTok{count}\NormalTok{(easy_dataset}\OperatorTok{$}\NormalTok{author)}
\NormalTok{comments <-}\StringTok{ }\NormalTok{comments}\OperatorTok{$}\NormalTok{freq}
\end{Highlighting}
\end{Shaded}

The command above simply counts the number of \emph{comments} by each
author of a social media comment. This number will be used later on to
generate a new variable \emph{comments}. In order to generate the number
of words and to generate a new variable \emph{words}, every sentence in
a comment must be broken down into its components, known as words. R
provides the \emph{strsplit(\ldots)} command in order to generate the
new variable \emph{words}:

\begin{Shaded}
\begin{Highlighting}[]
\NormalTok{temp <-}\StringTok{ }\KeywordTok{strsplit}\NormalTok{(easy_dataset}\OperatorTok{$}\NormalTok{sentence, }\DataTypeTok{split=}\StringTok{" "}\NormalTok{)}
\NormalTok{easy_dataset}\OperatorTok{$}\NormalTok{words <-}\StringTok{ }\KeywordTok{sapply}\NormalTok{(temp, length)}
\end{Highlighting}
\end{Shaded}

Only now and as one of the final steps of data pre-processing, the
double cases can be identified and have to be merged. This is necessary,
because some authors of social media comments have posted several
comments. Accordingly, the amounts of their numeric variables should be
summed up: If someone gets two likes for his first comment and three
likes for his second comment, he gets a total of five likes. This can be
done by using the \emph{ddply(\ldots)} command in the R terminal:

\begin{Shaded}
\begin{Highlighting}[]
\NormalTok{easy_dataset <-}\StringTok{ }\KeywordTok{ddply}\NormalTok{(easy_dataset, }\StringTok{"easy_dataset$author"}\NormalTok{, }\KeywordTok{numcolwise}\NormalTok{(sum))}
\end{Highlighting}
\end{Shaded}

Now it is possible to assign the correct number of comments for each
author of social media comments and to create a new variable
\emph{comments} inside the \emph{easy\_dataset}. This can be achieved by
applying the following command:

\begin{Shaded}
\begin{Highlighting}[]
\NormalTok{easy_dataset}\OperatorTok{$}\NormalTok{comments <-}\StringTok{ }\NormalTok{comments}
\KeywordTok{colnames}\NormalTok{(easy_dataset) <-}\StringTok{ }\KeywordTok{c}\NormalTok{(}\StringTok{"author"}\NormalTok{, }\StringTok{"replies"}\NormalTok{, }\StringTok{"likes"}\NormalTok{, }\StringTok{"words"}\NormalTok{, }\StringTok{"comments"}\NormalTok{)}
\end{Highlighting}
\end{Shaded}

This new and extended dataset, including two new generated numeric
variables, is suitable for further analyses. Since the variable
\emph{sentence} with all the qualitative data can not be summed up, it
is automatically excluded from the dataset by the previous commands.
However, it can be accessed any time before aggregating the double cases
and later on by using the initial dataset \emph{myYoutubeData}. Finally,
the \emph{head(\ldots)} command provides insights into this tidy
dataset, that still carries the designation \emph{easy\_dataset}. This
time only the quantitative variables will be focused, which can be
addressed by focusing the 2nd to 5th column of the dataset:

\begin{Shaded}
\begin{Highlighting}[]
\KeywordTok{head}\NormalTok{(easy_dataset[}\DecValTok{2}\OperatorTok{:}\DecValTok{5}\NormalTok{])}
\end{Highlighting}
\end{Shaded}

\begin{verbatim}
##   replies likes words comments
## 1       0     0    13        2
## 2       0     7    21        4
## 3       1     2    58        1
## 4       0     0     5        1
## 5       0     2    11        1
## 6       0     0    13        1
\end{verbatim}

The first six - and at this moment only disordered - cases can now
easily be inspected. For example, the second case reports \emph{0}
replies, \emph{7} likes, \emph{21} words and \emph{4} comments in total.

All collected 200 cases will be the basis for further steps in the
Analysis of Social Media Network Data. Within the next chapter, not only
the methodological approach for ordering the cases systematically and
identifying single authors of social media comments will be presented,
but also how to access their original comments in the previous dataset
for performing qualitative - and quantitative - analyses, if necessary.

\hypertarget{analysis---1-identification-of-cases-and-comments}{%
\subsection{Analysis - 1: Identification of Cases and
Comments}\label{analysis---1-identification-of-cases-and-comments}}

Using the dataset \emph{easy\_dataset} makes it possible to identify
particularly active users within the social media network. This is done
via the quantitative variables presented above. Since not all users of
social media networks get replies and likes, a first parameter for
identification could be a predefined minimum of at least one reply and
one like. This is probably the case, when they have written more than
one word, e.g.~ten words; Therefore they must have written at least one
comment. The identification parameters can be specified in the R
terminal as follows:

\begin{Shaded}
\begin{Highlighting}[]
\NormalTok{identification <-}\StringTok{ }\KeywordTok{subset}\NormalTok{(easy_dataset, replies}\OperatorTok{>=}\StringTok{"1"} \OperatorTok{&}\StringTok{ }\NormalTok{likes}\OperatorTok{>=}\StringTok{"1"}
  \OperatorTok{&}\StringTok{ }\NormalTok{words}\OperatorTok{>=}\StringTok{"10"} \OperatorTok{&}\StringTok{ }\NormalTok{comments}\OperatorTok{>=}\StringTok{"1"}\NormalTok{)}
\KeywordTok{head}\NormalTok{(identification)}
\end{Highlighting}
\end{Shaded}

\begin{verbatim}
##                 author replies likes words comments
## 3  anonymized username       1     2    58        1
## 10 anonymized username       1     6    36        1
## 13 anonymized username       1    15    19        1
## 16 anonymized username      10    58    58        4
## 17 anonymized username       2     6    31        2
## 19 anonymized username       3    14    48        1
\end{verbatim}

The result is a list of six particularly active users and their number
of replies, likes, words and comments. This provides a first insight
into the top-down communication patterns within the social media network
across several identification parameters. Since each user can be
addressed by his username (in this working paper anonymized) within the
variable \emph{author}, a user with a particular behavior can be
analysed individually. It is also possible to identify users by
addressing only one identification parameter:

\begin{Shaded}
\begin{Highlighting}[]
\NormalTok{most_replies <-}\StringTok{ }\NormalTok{easy_dataset[}\KeywordTok{order}\NormalTok{(}\OperatorTok{-}\NormalTok{easy_dataset}\OperatorTok{$}\NormalTok{replies),]}
\NormalTok{most_likes <-}\StringTok{ }\NormalTok{easy_dataset[}\KeywordTok{order}\NormalTok{(}\OperatorTok{-}\NormalTok{easy_dataset}\OperatorTok{$}\NormalTok{likes),]}
\NormalTok{most_words <-}\StringTok{ }\NormalTok{easy_dataset[}\KeywordTok{order}\NormalTok{(}\OperatorTok{-}\NormalTok{easy_dataset}\OperatorTok{$}\NormalTok{words),]}
\NormalTok{most_comments <-}\StringTok{ }\NormalTok{easy_dataset[}\KeywordTok{order}\NormalTok{(}\OperatorTok{-}\NormalTok{easy_dataset}\OperatorTok{$}\NormalTok{comments),]}
\end{Highlighting}
\end{Shaded}

With the presented command above, the dataset \emph{easy\_dataset} will
be ordered top-down, beginning with the maximum number of replies,
likes, words or comments. Each ranking can be addressed directly by its
designation, e.g.~\emph{most\_likes}. If someone is interested in the
user with the current maximum of likes, the username of this user will
be on top of that ranking:

\begin{Shaded}
\begin{Highlighting}[]
\KeywordTok{head}\NormalTok{(most_likes)}
\end{Highlighting}
\end{Shaded}

\begin{verbatim}
##                  author replies likes words comments
## 16  anonymized username      10    58    58        4
## 71  anonymized username       1    39     6        1
## 180 anonymized username       0    26     8        1
## 144 anonymized username       1    21     9        1
## 122 anonymized username       2    17    17        2
## 13  anonymized username       1    15    19        1
\end{verbatim}

Furthermore, the username within the variable \emph{author} provides
access to the sentences that were originally written by that user. This
can be done with recourse to the dataset \emph{myYoutubeData} and the
\emph{subset(\ldots)} command:

\begin{Shaded}
\begin{Highlighting}[]
\NormalTok{sentences <-}\StringTok{ }\KeywordTok{subset}\NormalTok{(myYoutubeData[}\KeywordTok{c}\NormalTok{(}\DecValTok{1}\OperatorTok{:}\DecValTok{2}\NormalTok{)], AuthorDisplayName}\OperatorTok{==}\StringTok{"ENTER USERNAME HERE"}\NormalTok{)}
\NormalTok{sentences[}\KeywordTok{c}\NormalTok{(}\DecValTok{1}\NormalTok{), }\DecValTok{1}\NormalTok{]}
\end{Highlighting}
\end{Shaded}

Therefore, this example will focus the 16th case, again displayed as
``anonymized username'', since all usernames have been anonymized for
this working paper. This user not only got the highest number on likes,
but also wrote more comments than other users and collected several
replies. Therefore, his example clarifies the output in R for each
single sentence, which can be directly addressed by changing the second
number within the command. Hence, this is the first sentence of this
specific user:

\begin{verbatim}
## # A tibble: 1 x 1
##   Comment                                        
##   <chr>                                          
## 1 Being a sore loser isn't grounds for a lawsuit.
\end{verbatim}

By accessing the sentences, there are a lot of things that can be
analysed from here one. Sentiment analysis and semantic analysis of
sentences are only two examples. Furthermore a simple word cloud could
be generated by using R, in order to focus on keywords used within the
social media network.

However, this techniques will be highlighted in a separate paper,
because after this demonstration on how individual cases and comments
can be identified and addressed among several hundreds or thousands of
cases. How easy it can be to identify users as conversation partners of
a specific user will be highlighted in the chapter about the
visualization of social networks. Next aim is to gain a general
understanding of the communication patterns within specific contents on
social media platforms.

\hypertarget{analysis---2-identification-of-communication-patterns}{%
\subsection{Analysis - 2: Identification of Communication
Patterns}\label{analysis---2-identification-of-communication-patterns}}

A social network is characterized by active and less active users. In
particular, active users tend to shape the communication patterns of
social networks more likely than their lesser-active counterparts. In
theory, some of these active users seem to reach out to others in order
to share their opinion; That is what makes them relevant for Analysing
Social Media Network Data. Therefore, the above-average active users and
the below-average active users are identified by using the
\emph{summary(\ldots)} command first:

\begin{Shaded}
\begin{Highlighting}[]
\KeywordTok{summary}\NormalTok{(easy_dataset[}\KeywordTok{c}\NormalTok{(}\DecValTok{2}\OperatorTok{:}\DecValTok{5}\NormalTok{)])}
\end{Highlighting}
\end{Shaded}

In the following output, the mean values will be used to differentiate
between above-average active users and below-average active users:

\begin{verbatim}
##     replies            likes            words           comments    
##  Min.   : 0.0000   Min.   : 0.000   Min.   :  1.00   Min.   :1.000  
##  1st Qu.: 0.0000   1st Qu.: 0.000   1st Qu.:  6.00   1st Qu.:1.000  
##  Median : 0.0000   Median : 0.000   Median : 12.00   Median :1.000  
##  Mean   : 0.2574   Mean   : 1.748   Mean   : 27.12   Mean   :1.248  
##  3rd Qu.: 0.0000   3rd Qu.: 1.000   3rd Qu.: 25.75   3rd Qu.:1.000  
##  Max.   :10.0000   Max.   :58.000   Max.   :561.00   Max.   :7.000
\end{verbatim}

The mean values can then be implemented in the \emph{ifelse(\ldots)}
command in the R terminal in order to create a new variable
\emph{relevant}.

\begin{Shaded}
\begin{Highlighting}[]
\NormalTok{easy_dataset}\OperatorTok{$}\NormalTok{relevant <-}\StringTok{ }\KeywordTok{ifelse}\NormalTok{(easy_dataset}\OperatorTok{$}\NormalTok{replies}\OperatorTok{>=}\FloatTok{0.25} \OperatorTok{&}\StringTok{ }\NormalTok{easy_dataset}\OperatorTok{$}\NormalTok{likes}\OperatorTok{>=}\FloatTok{1.74}
  \OperatorTok{&}\StringTok{ }\NormalTok{easy_dataset}\OperatorTok{$}\NormalTok{words}\OperatorTok{>=}\FloatTok{27.12} \OperatorTok{&}\StringTok{ }\NormalTok{easy_dataset}\OperatorTok{$}\NormalTok{comments}\OperatorTok{>=}\FloatTok{1.24}\NormalTok{, }\DecValTok{1}\NormalTok{,}\DecValTok{0}\NormalTok{)}
\end{Highlighting}
\end{Shaded}

Again, the result can be inspected by using the \emph{summary(\ldots)}
command:

\begin{Shaded}
\begin{Highlighting}[]
\KeywordTok{summary}\NormalTok{(easy_dataset[}\KeywordTok{c}\NormalTok{(}\DecValTok{6}\NormalTok{)])}
\end{Highlighting}
\end{Shaded}

Presented below is the output of the new variable \emph{relevant}, which
can be found in the six column of the dataset. This variable has been
generated as binary variable, which means that above-average active
users can be identified as \emph{1} and below-average active users as
\emph{0}:

\begin{verbatim}
##     relevant     
##  Min.   :0.0000  
##  1st Qu.:0.0000  
##  Median :0.0000  
##  Mean   :0.0198  
##  3rd Qu.:0.0000  
##  Max.   :1.0000
\end{verbatim}

It is also possible to convert the remaining variables into binary
variables. This has the advantage that all variables have a uniform
scale level. Based on a uniform scale level, the next step is to set up
an analysis model to identify the communication patterns. First,
however, the command for converting the predictors into binary variables
will be presented:

\begin{Shaded}
\begin{Highlighting}[]
\NormalTok{easy_dataset}\OperatorTok{$}\NormalTok{replies <-}\StringTok{ }\KeywordTok{ifelse}\NormalTok{(easy_dataset}\OperatorTok{$}\NormalTok{replies}\OperatorTok{>=}\FloatTok{0.25}\NormalTok{,}\DecValTok{1}\NormalTok{,}\DecValTok{0}\NormalTok{)}
\NormalTok{easy_dataset}\OperatorTok{$}\NormalTok{likes <-}\StringTok{ }\KeywordTok{ifelse}\NormalTok{(easy_dataset}\OperatorTok{$}\NormalTok{likes}\OperatorTok{>=}\FloatTok{1.74}\NormalTok{,}\DecValTok{1}\NormalTok{,}\DecValTok{0}\NormalTok{)}
\NormalTok{easy_dataset}\OperatorTok{$}\NormalTok{words <-}\StringTok{ }\KeywordTok{ifelse}\NormalTok{(easy_dataset}\OperatorTok{$}\NormalTok{words}\OperatorTok{>=}\FloatTok{27.12}\NormalTok{,}\DecValTok{1}\NormalTok{,}\DecValTok{0}\NormalTok{)}
\NormalTok{easy_dataset}\OperatorTok{$}\NormalTok{comments <-}\StringTok{ }\KeywordTok{ifelse}\NormalTok{(easy_dataset}\OperatorTok{$}\NormalTok{comments}\OperatorTok{>}\FloatTok{1.24}\NormalTok{,}\DecValTok{1}\NormalTok{,}\DecValTok{0}\NormalTok{)}
\end{Highlighting}
\end{Shaded}

These binary variables can also be inspected via the
\emph{summary(\ldots)} command in the R terminal:

\begin{Shaded}
\begin{Highlighting}[]
\KeywordTok{summary}\NormalTok{(easy_dataset[}\KeywordTok{c}\NormalTok{(}\DecValTok{2}\OperatorTok{:}\DecValTok{5}\NormalTok{)])}
\end{Highlighting}
\end{Shaded}

As a result, all predictors vary between a minimum of zero and a maximum
of one. The predictors can now be used to predict whether a user shows
relevant behavior inside a social network, or not:

\begin{verbatim}
##     replies           likes           words           comments     
##  Min.   :0.0000   Min.   :0.000   Min.   :0.0000   Min.   :0.0000  
##  1st Qu.:0.0000   1st Qu.:0.000   1st Qu.:0.0000   1st Qu.:0.0000  
##  Median :0.0000   Median :0.000   Median :0.0000   Median :0.0000  
##  Mean   :0.1386   Mean   :0.203   Mean   :0.2327   Mean   :0.1634  
##  3rd Qu.:0.0000   3rd Qu.:0.000   3rd Qu.:0.0000   3rd Qu.:0.0000  
##  Max.   :1.0000   Max.   :1.000   Max.   :1.0000   Max.   :1.0000
\end{verbatim}

An adequate model would examine whether users who have the value
\emph{1} for the predictors also have a value of \emph{1} for the
variable \emph{relevant}. Then the weight of the predictors can be
calculated, which are representative for the communication patterns of
any specific social network. A model using binary variables can be
called with the \emph{glm (\ldots)} command with reference to
\emph{family=binomial()} in the R terminal:

\begin{Shaded}
\begin{Highlighting}[]
\NormalTok{glm.fit <-}\StringTok{ }\KeywordTok{glm}\NormalTok{(}\DataTypeTok{data=}\NormalTok{easy_dataset, relevant}\OperatorTok{~}\NormalTok{replies}\OperatorTok{+}\NormalTok{likes}\OperatorTok{+}\NormalTok{words}\OperatorTok{+}\NormalTok{comments,}
  \DataTypeTok{family=}\KeywordTok{binomial}\NormalTok{())}
\NormalTok{glm.fit}
\end{Highlighting}
\end{Shaded}

A logistic regression model with the weights of the predictors is shown
in the output. The weights can be found directly below the predictors,
indicated by the names of the variables, and they vary between a minimum
weight of \emph{41.44} and a maximum weight of \emph{42.61}. The
procedure using such a model is also often referred to as Artificial
Neural Network (ANN), however, in this instance only on a very simple
basis:

\begin{verbatim}
## 
## Call:  glm(formula = relevant ~ replies + likes + words + comments, 
##     family = binomial(), data = easy_dataset)
## 
## Coefficients:
## (Intercept)      replies        likes        words     comments  
##     -147.84        42.51        41.44        42.27        42.61  
## 
## Degrees of Freedom: 201 Total (i.e. Null);  197 Residual
## Null Deviance:       39.3 
## Residual Deviance: 1.622e-08     AIC: 10
\end{verbatim}

This slight difference can be considered a ranking of all variables in
order to describe the communication pattern whithin this specific social
network. Therefore, the variable \emph{comments} is more important for
being an above-average active user than the variables \emph{replies},
\emph{words} and \emph{likes} (in descending order). These results tend
to vary, depending on the size of the network, the activity of the users
as well as the content of the access point. Therefore, social networks
referring to political content have different communication patterns
than those referring to sports, music or other contents, as will be
highlighted later on via graphical representation.

Instead of using a uniform method for analysing users of social media
platforms, a content- and network-specific approach is recommended. Only
an interplay of sensitivity and specificity enables the identification
of particularly active - and relevant - users, their comments and unique
patterns via the predictors used in this statistical analysis. This
makes it possible to identify opinion leaders, polarizing content and
heated debates more quickly, compared to manual inspections of all
users. In addition, one can see whether the users stand out in
particular because they write a lot of comments, use a lot of words, get
a lot of likes or how often they receive replies.

If necessary, a list of identified cases can be called by using the
\emph{which(\ldots)} command in R, subsequent after computing the
predictions of the logistic regression model with the
\emph{predict(\ldots)} command and assigning the specification
parameters \emph{0} and \emph{1} within the \emph{ifelse(\ldots)}
command. In this instance, cases \emph{16}, \emph{17}, \emph{30} and
\emph{35} can be identified as relevant users that frame a hot spot of
communication within the social network. In addition, the predictions
can be compared with the actual values of the variable \emph{relevant},
which in this case results in an accuracy of 100 percent. The complete
list of required commands is presented below:

\begin{Shaded}
\begin{Highlighting}[]
\NormalTok{probabilities <-}\StringTok{ }\NormalTok{glm.fit }\OperatorTok{%>%}\StringTok{ }\KeywordTok{predict}\NormalTok{(easy_dataset, }\DataTypeTok{type =} \StringTok{"response"}\NormalTok{)}
\NormalTok{predicted.classes <-}\StringTok{ }\KeywordTok{ifelse}\NormalTok{(probabilities }\OperatorTok{>}\StringTok{ }\FloatTok{0.5}\NormalTok{, }\StringTok{"1"}\NormalTok{, }\StringTok{"0"}\NormalTok{)}
\KeywordTok{which}\NormalTok{(predicted.classes}\OperatorTok{==}\StringTok{"1"}\NormalTok{)}
\KeywordTok{mean}\NormalTok{(predicted.classes }\OperatorTok{==}\StringTok{ }\NormalTok{easy_dataset}\OperatorTok{$}\NormalTok{relevant)}
\end{Highlighting}
\end{Shaded}

In contrast to what is shown in this example, by using mainly mean
values, the relevant cases can also be designated manually on basis of
qualitative inspection of the comments. With this method, for example,
the communication patterns of polarizing and appeasing users could
easily be compared with one another, even if it can be assumed that the
accuracy will then be slightly lower. This example, however, has focused
mainly the above-average users for demonstration purpose only.

\hypertarget{analysis---3-visualization-of-social-networks}{%
\subsection{Analysis - 3: Visualization of Social
Networks}\label{analysis---3-visualization-of-social-networks}}

The identification of cases and comments as well as a basic
understanding of communication patterns on social media platforms are
necessary for visualizing social networks. Thereby, relevant users as
well as individual topics and the dissemination of these topics can be
made clear throughout the comments and replies of a social network. In
order to do so, the initial dataset \emph{myYoutubeData} will be
transformed into a dataset ready for visualization that focuses the
activity within the network via the \emph{Create(\ldots)} command.
Activity networks are based upon the comments, represented by nodes
within each network and edges represent the direction of these comments,
either to the social media content itself or to other comments. The
result is a Graph-object, stored as \emph{activityGraph}:

\begin{Shaded}
\begin{Highlighting}[]
\NormalTok{activityNetwork <-}\StringTok{ }\NormalTok{myYoutubeData }\OperatorTok{%>%}\StringTok{ }\KeywordTok{Create}\NormalTok{(}\StringTok{"activity"}\NormalTok{) }\OperatorTok{%>%}\StringTok{ }\KeywordTok{AddText}\NormalTok{(myYoutubeData)}
\NormalTok{activityGraph <-}\StringTok{ }\NormalTok{activityNetwork }\OperatorTok{%>%}\StringTok{ }\KeywordTok{Graph}\NormalTok{(}\DataTypeTok{writeToFile =} \OtherTok{TRUE}\NormalTok{)}
\end{Highlighting}
\end{Shaded}

Since Graph-objects as well as the underlying dataset on the activity
network differ from common datasets in terms of datastructure, the
package \emph{iGraph} is needed - which, at this step in Analysing
Social Media Network Data, should be already installed via the previous
commands - in order to process them:

\begin{Shaded}
\begin{Highlighting}[]
\KeywordTok{library}\NormalTok{(igraph)}
\end{Highlighting}
\end{Shaded}

The difference to a normal dataset leads to a different way of writing
the commands. For instance, the names of the authors of the comments are
now stored inside the sixth column of the variable \emph{nodes} of the
\emph{activityNetwork}, whereas in a common dataset the columns
represent only one variable. If one wants to put the focus back on the
already identified user ``anonymized username'', his name needs to be
added to the following command:

\begin{Shaded}
\begin{Highlighting}[]
\KeywordTok{which}\NormalTok{(activityNetwork}\OperatorTok{$}\NormalTok{nodes[}\DecValTok{6}\NormalTok{]}\OperatorTok{==}\StringTok{"ENTER USERNAME HERE"}\NormalTok{)}
\end{Highlighting}
\end{Shaded}

\begin{verbatim}
## [1] 134 230 235 236
\end{verbatim}

The result is the specific number of cases the user commented on
YouTube. Since within a Graph-object the number of cases are the
comments and not the authors of the comments, this specific number is
needed in order to identify users afterwards within the activity
network. Any user can be identified within an activity network by his
\emph{author\_id}, which is usually represented by a random and
anonymized number, but the specific case number enables direct access to
this random and anonymized number:

\begin{Shaded}
\begin{Highlighting}[]
\NormalTok{(}\KeywordTok{V}\NormalTok{(activityGraph)}\OperatorTok{$}\NormalTok{author_id[ENTER CASE NUMBER HERE])}
\end{Highlighting}
\end{Shaded}

\begin{verbatim}
## [1] "UCtICekaZ9ptKjKHPBJxBSbA"
\end{verbatim}

It is also possible to reverse this technique in order to identify the
conversation partners of a specific user by their usernames. To identify
recipients of a comment one has to subtract 1 from the particular case
number of interest and in order to identify replies to that comment one
can add a consecutive series of numbers, starting with 1. This is due to
the fact that a comment can only be addressed to one other user, but
several users can refer to one comment. A corresponding command would
look like this in R:

\begin{Shaded}
\begin{Highlighting}[]
\NormalTok{activityNetwork}\OperatorTok{$}\NormalTok{nodes[ENTER CASE NUMBER HERE, }\DecValTok{6}\NormalTok{]}
\end{Highlighting}
\end{Shaded}

This could also be used to call up the comments of these users - as
previously shown - and their positions as conversation partners in the
social network. Therefore it is possible to highlight relevant elements
in color:

\begin{center}\includegraphics{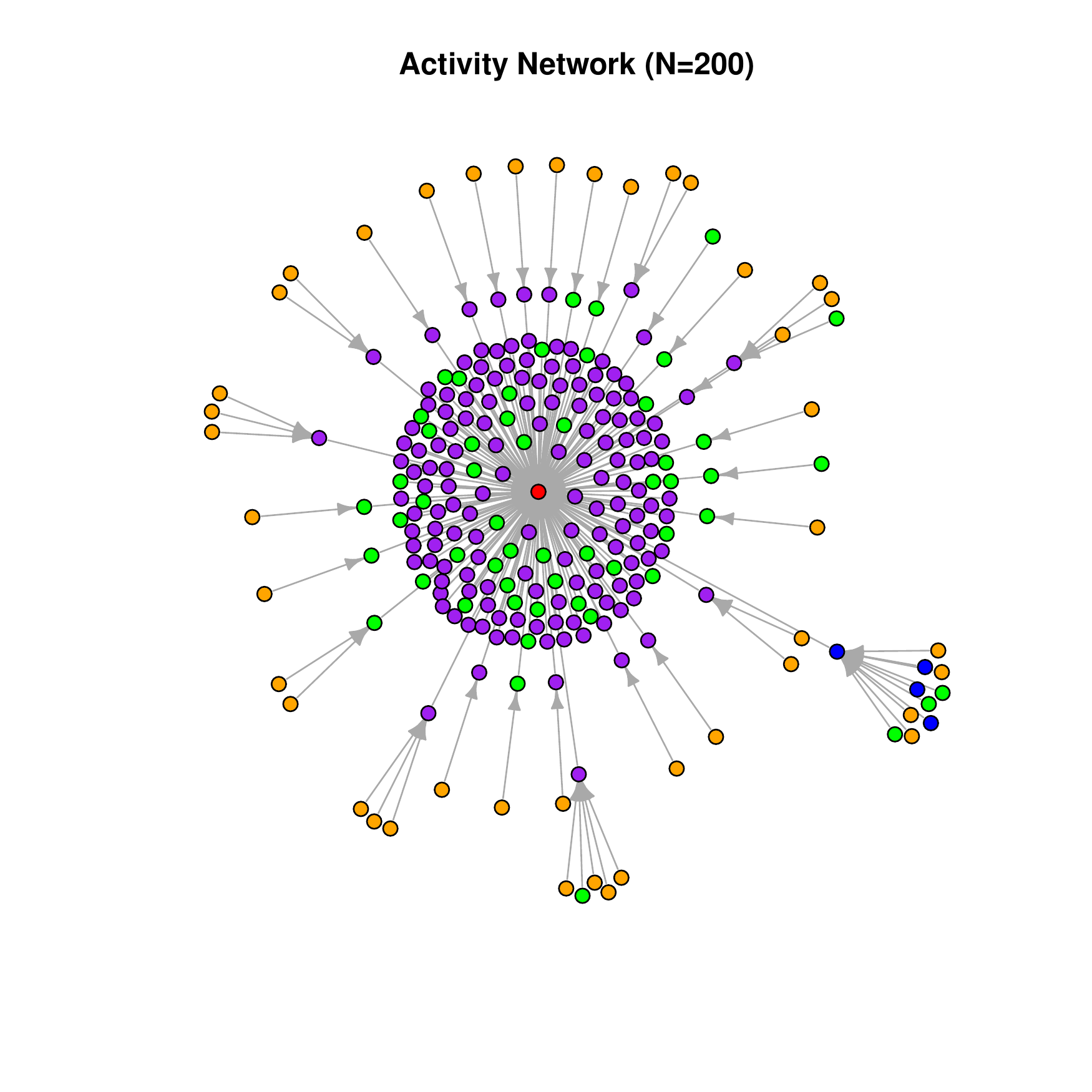} \end{center}

For example, the video itself is highlighted in red, each comment is
highlighted in purple, focusing the 16th case can be done by referring
to his random and anonymized \emph{author\_id} and assigning him blue as
color and it is also possible to highlight each comment that contains a
predefined list of buzzwords, such as ``Donald Trump'', ``President'' or
``White House''. One can add any number of buzzwords that either have to
occur together, represented by a \emph{\&} sign, or can also occur
individually, represented by a \emph{\textbar{}} sign. As a result, the
distribution of these buzzwords can be visualized throughout the entire
social network. All remaining comments are replies and will be plotted
in orange, as assigned by the following command:

\begin{Shaded}
\begin{Highlighting}[]
\KeywordTok{V}\NormalTok{(activityGraph)}\OperatorTok{$}\NormalTok{color <-}\StringTok{ "orange"}
\KeywordTok{V}\NormalTok{(activityGraph)}\OperatorTok{$}\NormalTok{color[}\KeywordTok{which}\NormalTok{(}\KeywordTok{V}\NormalTok{(activityGraph)}
  \OperatorTok{$}\NormalTok{node_type}\OperatorTok{==}\StringTok{"video"}\NormalTok{)] <-}\StringTok{ "red"}
\KeywordTok{V}\NormalTok{(activityGraph)}\OperatorTok{$}\NormalTok{color[}\KeywordTok{which}\NormalTok{(}\KeywordTok{V}\NormalTok{(activityGraph)}
  \OperatorTok{$}\NormalTok{node_type}\OperatorTok{==}\StringTok{"comment"}\NormalTok{)] <-}\StringTok{ "purple"}
\KeywordTok{V}\NormalTok{(activityGraph)}\OperatorTok{$}\NormalTok{color[}\KeywordTok{which}\NormalTok{(}\KeywordTok{V}\NormalTok{(activityGraph)}
  \OperatorTok{$}\NormalTok{author_id}\OperatorTok{==}\StringTok{"ENTER AUTHOR ID HERE"}\NormalTok{)] <-}\StringTok{ "blue"}
\NormalTok{marker <-}\StringTok{ }\KeywordTok{grep}\NormalTok{(}\StringTok{"ENTER BUZZWORDS HERE"}\NormalTok{,}
  \KeywordTok{tolower}\NormalTok{(}\KeywordTok{V}\NormalTok{(activityGraph)}\OperatorTok{$}\NormalTok{vosonTxt_comment))}
\KeywordTok{V}\NormalTok{(activityGraph)}\OperatorTok{$}\NormalTok{color[marker] <-}\StringTok{ "green"}
\end{Highlighting}
\end{Shaded}

The standard \emph{plot(\ldots)} command of R can also be specified in
terms of the sizes of each element within the plot, but the setting
presented below should enable a readable plot for most cases, as
presented before:

\begin{Shaded}
\begin{Highlighting}[]
\KeywordTok{plot}\NormalTok{(activityGraph, }\DataTypeTok{vertex.label=}\StringTok{""}\NormalTok{, }\DataTypeTok{vertex.size=}\DecValTok{4}\NormalTok{,}
  \DataTypeTok{edge.arrow.size=}\FloatTok{0.6}\NormalTok{, }\DataTypeTok{main=}\StringTok{"Activity Network (N=200)"}\NormalTok{)}
\end{Highlighting}
\end{Shaded}

In summary, this activity network visualizes the Youtube video in the
center of the social network, surrounded by a number of comments
directed towards it. These comments are thereby initial comments.
Comments referring to the initial comments can be identified by a larger
distance to the video and by referring to other comments and not to the
video itself. Regardless of their proximity to the center of the social
network, ordinary comments are shown in purple, as predefined before.
Non-ordinary comments, e.g.~when containing some of the predefined
buzzwords, are highlighted in green. And the comments of the 16th case
with an ``anonymized username'' are plotted in blue, so that it is
possible to assess the role of specific users and buzzwords when it
comes to the dynamics within social networks. Further application
examples follow in the next section, in order to highlight the different
types of dynamics by the topic of a social network.

\hypertarget{application-examples-and-comparison-of-communication-patterns}{%
\subsection{Application Examples and Comparison of Communication
Patterns}\label{application-examples-and-comparison-of-communication-patterns}}

In a final step, four application examples are to be compared with one
another: Music, sports, politics and games. All data is collected on
YouTube via specific access points that are representative for all four
different topics, are frequently watched on YouTube and strongly
commented on.

As a first result, comparable topics, with the same number of comments
to be analysed, tend to generate social networks with a similar
structure when it comes to the distribution of comments and replies, as
well as the scattering of buzzwords throughout the social network.
Buzzwords, for example, are not only more often used when talking about
sports and politics, they also seem to have a greater range in these
social networks. The range can be addressed by counting the number or
outer rings of the social network, which represent replies to previous
replies or comments. Within these social networks, the topics seem to
run through several levels of discussion, contrary to the social
networks regarding music or games.

\begin{center}\includegraphics{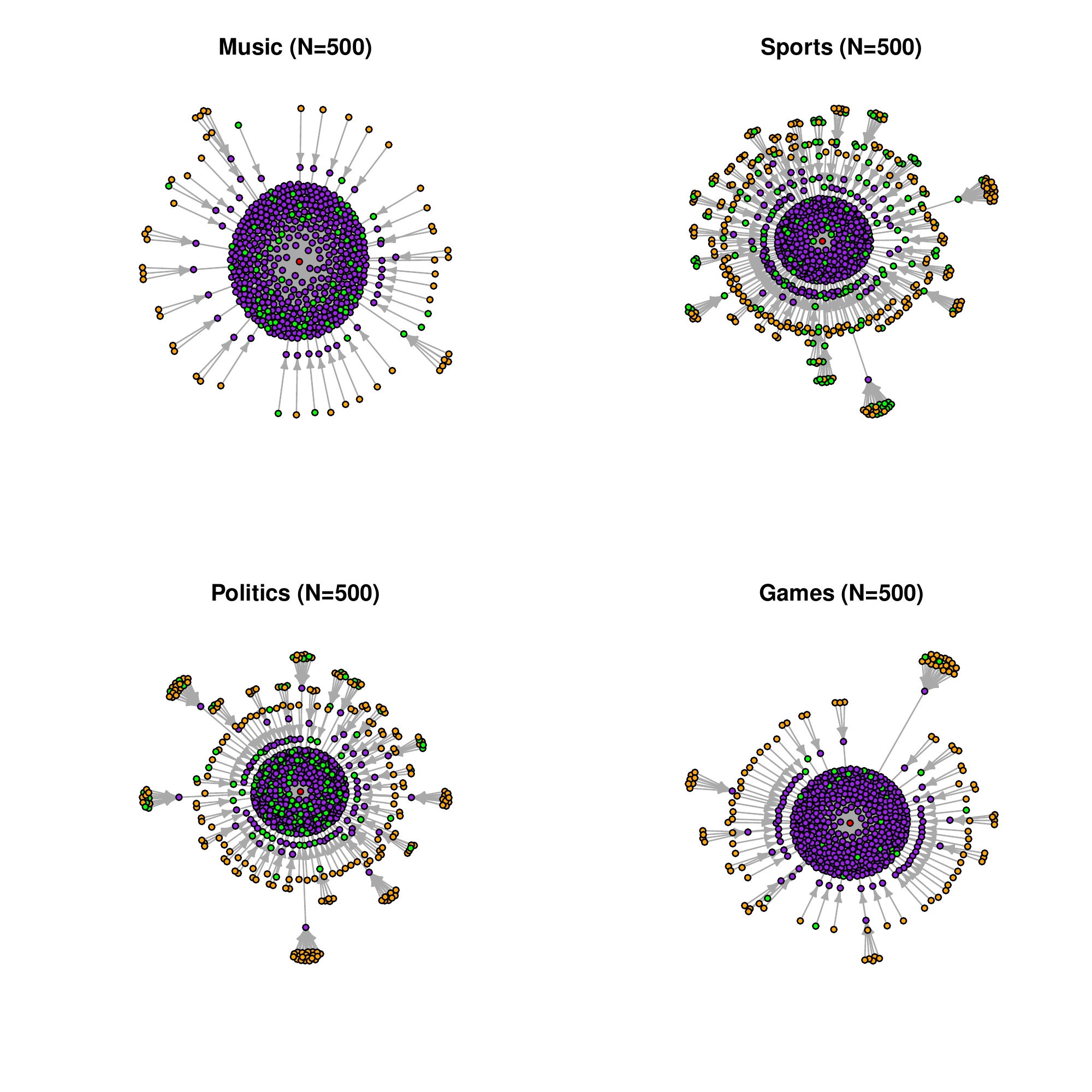} \end{center}

As a second result, further variables could be generated, based on the
properties of each social network (e.g.~by counting the appearances of
buzzwords or the number of outer rings). This not only makes it possible
to identify individual users and their specific role within a social
network more sensitive, but to increase the specificity of the search
for relevant users by taking these different types of social network
structures into account. Furthermore, the scattering of buzzwords can be
compared on an inter-topic level and be combined with sentiment
analysis, semantic analysis and qualitative content analysis by
identifying the position of each comment within a social network and the
reactions it created.

\hypertarget{closing-remarks}{%
\subsection{Closing Remarks}\label{closing-remarks}}

A methodological approach based on the R programming language was
presented, so that social media platforms like YouTube, Twitter or
Reddit can be analysed in few and reproducible steps.

In a first step, an initial dataset based upon the social media platform
YouTube has been converted into accessible variables via \emph{vosonSML}
and pre-processed in order to generate a dataset which is easy to
understand, in a few steps to modify and comprehensible to analyse. In a
second step, it was presented how further quantitative variables can be
generated on the basis of qualitative variables. This setting of
quantitative variables has been used in order to identify particularly
active users within a social network.

Furthermore, as a third step, since the comments of particularly active
users could contain relevant content, e.g.~from a scientific point of
view or in order to identify fake news, hate speech and radicalizing
elements, a method was shown how to access these comments among an often
confusing large number of other comments, so that they can be inspected
individually or compared among others. This assessment is subject to the
scientists or analysing persons themselves.

After the content has been classified as relevant, the fourth step broke
down the underlying communication patterns. As a result, the same users
of social media platforms could be highlighted in the output, that
already have been identified within the third step. A method for
classification using the mean values of the variables was presented in
order to do so; The accuracy was 100 percent. Other approaches are also
conceivable, provided they can be theoretically justified and tested
afterwards. The overlap between identified users in these two steps
confirm this procedure. Based on the communication patterns, indicated
mainly by focusing the weight of the variables \emph{likes},
\emph{replies}, \emph{words} and \emph{comments}, the impact and the
systematics of active users within a social network could be pinpointed.

It must be taken into account, that every social network, depending on
its users and topics, can have its own dynamics and thus different
weights for these variables. This makes uniform identification
approaches more difficult, if not impossible. The influence of these
dynamics, which seems to emanate mainly from the topics discussed within
a social network, can also be highlighted graphically. Examples have
been provided by focusing social networks on music, sports, politics and
gaming as primary topics. All social networks differ in terms of
density, their range and the distribution of the topic within the
comments. However, different social networks on a similar topic are more
alike than social networks with a different topic.

This indicates that the general communication pattern should first be
identified before the particularities of individual users can be
highlighted. With this temporary result, the working paper, which is
also a tutorial on the R programming language in terms of Analysing
Social Media Network Data, provides a basis for further research and a
methodological approach for several scientific disciplines. All codes
are also available on GitHub.

\hypertarget{sources}{%
\subsection{Sources}\label{sources}}

\begin{itemize}
\item
  Awan, Imran (2017): ``Cyber-Extremism: Isis and the Power of Social
  Media''. Online:
  \url{https://www.researchgate.net/publication/315212548_Cyber-Extremism_Isis_and_the_Power_of_Social_Media}
\item
  Bache, Stefan Milton \& Hadley Wickham (2020): ``magrittr: A
  Forward-Pipe Operator for R''. Online:
  \url{https://cran.r-project.org/package=magrittr}
\item
  Cement, Jessica (2020): ``Most popular social networks worldwide''.
  Online:
  \url{https://www.statista.com/statistics/272014/global-social-networks-ranked-by-number-of-users/}
\item
  Fielding, Roy Thomas (2000): Architectural Styles and the Design of
  Network-based Software Architectures". Online:
  \url{https://www.ics.uci.edu/~fielding/pubs/dissertation/top.htm}
\item
  Graham, Timothy; Ackland, Robert; Chan, Chung-hong \& Bryan Gertzel
  (2020): ``vosonSML: Collecting Social Media Data and Generating
  Networks for Analysis''. Online:
  \url{https://cran.r-project.org/package=vosonSML}
\item
  Machackova, Hana; Blaya, Catherine; Bedrosova, Marie; Smahel, David \&
  Elisabeth Staksrud (2020): ``Children's experiences with cyberhate''.
  Online:
  \url{https://www.lse.ac.uk/media-and-communications/assets/documents/research/eu-kids-online/reports/euko-cyberhate-22-4-final.pdf}
\item
  Ott, Brian (2016): ``The age of Twitter: Donald J. Trump and the
  politics ofdebasement''. Online:
  \url{https://www.researchgate.net/publication/311892973_The_age_of_Twitter_Donald_J_Trump_and_the_politics_of_debasement}
\item
  Pereira-Kohatsu, Juan Carlos; Quijano-Sánchez, Lara; Liberatore,
  Federico \& Miguel Camacho-Collados (2019): ``Detecting and Monitoring
  Hate Speech in Twitter''. Online:
  \url{https://www.mdpi.com/1424-8220/19/21/4654}
\item
  Reichelmann, Ashley; Hawdon, James; Costello, Matt; Ryan, John; Blaya,
  Catherine; Llorent, Vincente; Oksanen, Atte; Räsänen, Pekka \& Izabela
  Zych (2020): ``Hate Knows No Boundaries: Online Hate in Six Nations''.
  Online:
  \url{https://www.researchgate.net/publication/338899829_Hate_Knows_No_Boundaries_Online_Hate_in_Six_Nations}
\item
  Wickham, Hadley (2019): ``stringr: Simple, Consistent Wrappers for
  Common String Operations''. Online:
  \url{https://cran.r-project.org/package=stringr}
\item
  Wickham, Hadley (2020): ``plyr: Tools for Splitting, Applying and
  Combining Data''. Online:
  \url{https://cran.r-project.org/package=plyr}
\end{itemize}

\hypertarget{author-affiliations-and-materials-on-github}{%
\subsection{Author, Affiliations and Materials on
GitHub}\label{author-affiliations-and-materials-on-github}}

Dennis Klinkhammer is Professor for Empirical Research at the FOM
University of Applied Sciences. He advises public as well as
governmental organisations on the application of multivariate statistics
and limitations of artificial intelligence by providing introductions to
Python and R: \url{https://www.statistical-thinking.de}

All codes required for an Analysis of Social Media Network Data with R
can be accessed on GitHub:
\url{https://github.com/statistical-thinking/social-media-analysis}

\end{document}